\documentclass[preprint]{revtex4} 
\usepackage{mathrsfs}
\usepackage{bm}
\usepackage{amsmath}

\begin{document}
\title{General Properties of Two-dimensional Conformal Transformation in
Electrostatics}
\author{Yong Zeng,$^{1}$ Jinjie Liu,$^{2}$ and Douglas H. Werner$^{1}$}
\address{$^1$Department of Electrical
Engineering, Pennsylvania State University, University Park, PA
16802\\ $^2$Department of Mathematical Sciences, Delaware State
University, Dover, DE 19901}
\input epsf
\begin{abstract}
Electrostatic properties of two-dimensional nanosystems can be
described by their geometry resonances. In this paper we prove
that these modes as well as the corresponding eigenvalues are
invariant under any conformal transformation. This invariance
further leads to a new way to studying the transformed structures.
A special kind of three-dimensional transformations are further
investigated.
\end{abstract}
\maketitle

One specular property of Maxwell's equations is that their form
will be invariant under arbitrary coordinate transformation if the
field quantities as well as the material properties are
transformed accordingly \cite{pendry,leonhardt1,leonhardt2}, which
leads to a powerful designing tool called transformation optics or
transformation electromagnetism in 2006 \cite{pendry,leonhardt1}.
Since then, a wealth of novel and unique devices have been
theoretically suggested and/or experimentally fabricated, see
recent reviews \cite{kwon,chen} and the references cited. An
incomplete list includes electromagnetic cloaks
\cite{pendry,leonhardt1,schurig,liu,Ergin}, event cloaks
\cite{mcCall}, optical black holes \cite{Narimanov} and field
splitters \cite{rahm}.

Quasi-static approximations can be applied when electromagnetic
wavelength is far longer than the characteristic size of the
material. The interactions between light and medium hence can be
approximately described by Laplace's equation. Similar to the
full-wave Maxwell equations, two-dimensional Laplacian equation is
invariant under conformal transformations \cite{jackson}.
Furthermore, the electrostatic transformation does not alter the
properties of the constituent material, in sharp contrast to the
full-wave coordinate transformations which generally transform
normal medium to exotic and strange one. Lately, this unique
property of Laplacian equation is used to study two-dimensional
plasmonic particles
\cite{aubry1,aubry2,aubry3,aubry4,aubry5,aubry6}. More
specifically, two-dimensional complicated structures, by
transforming to simple one-dimensional geometries such as
finite-thickness metallic slab, can be studied analytically
\cite{aubry1,aubry2,aubry3,aubry4,aubry5,aubry6}. It is further
found that these plasmonic nanoparticles, include crescent and
touching or non-touching cylinder dimers, can be broadband and
significantly enhance the local electric fields.

In this paper, we will study electrostatic conformal
transformation in a new way, by taking advantage of the invariance
of geometry resonances and their eigenvalues. We will show that
transforming geometry can be equivalently explained as modifying
the amplitudes of these geometry resonances by transforming the
excitation source. Furthermore, a few energy quantities such as
electrostatic energy possessed by the particle are found to be
conserved under conformal transformation. Decreasing the area of
the particle will then definitely increasing the degree of local
field enhancement.

We briefly recall the spectral Bergman-Milton theory
\cite{bergman,milton}. It is stated that the electrostatic
behavior of a two-constituent composite can be described by a set
of eigenmodes $\varphi_{n}$ of the following generalized
eigenproblem
\begin{equation}
\nabla\cdot\left[\theta(\mathbf{r})\nabla\varphi_{n}(\mathbf{r})\right]=
s_{n}\nabla^{2}\varphi_{n}(\mathbf{r}), \label{eq1}
\end{equation}
where $s_{n}$ representing the corresponding eigenvalues, and the
function $\theta(\mathbf{r})$ characterizing the geometry of the
composite, equal to 1 inside one constituent, with a permittivity
of $\epsilon_{1}(\omega)$, and 0 inside the other medium with a
permittivity of $\epsilon_{2}(\omega)$ (we assume $\epsilon_{2}=1$
in the following to simplify the discussion). Since this equation
depends exclusively on the geometry, but not on the material
composition, the resultant eigenmodes are therefore named as
geometry resonances \cite{bergman}. Moreover, by defining a scalar
product as
\begin{equation}
(\phi|\psi)=\int\nabla\phi^{\ast}\cdot\nabla\psi\:dv, \label{eq2}
\end{equation}
we will find that $s_{n}$ must be real and limited as $0\leq
s_{n}\leq 1$, and the normalized eigenmodes $\varphi_{n}$ form a
complete orthonormal set \cite{bergman}. Note that this theory has
been successfully employed to predict spaser \cite{bergman2} and
study ultrafast optical excitation in nanosystems
\cite{mark1,mark2}.

It is important to notice, for a two-dimensional system, the
eigenequation (\ref{eq1}) remains invariance under any conformal
coordinate transformation. To prove this fact, we rewrite the
left-hand side of equation (\ref{eq1}) as
\begin{equation}
\partial_{i}\mathbf{e}_{i}\cdot(\theta\partial_{j}\varphi_{n}\mathbf{e}_{j})=g_{ij}\partial_{i}(\theta\partial_{j}\varphi_{n})=g\partial_{i}(\theta\partial_{i}\varphi_{n}),
\label{eq3}
\end{equation}
where the Einstein summation convention is employed, and the
metric tensor
$g_{ij}\equiv\mathbf{e}_{i}\cdot\mathbf{e}_{j}=g\delta_{ij}$
because of the conformal transformation \cite{leonhardt1}.
Similarly the right-hand side can be reformulated as
\begin{equation}
s_{n}\partial_{i}\mathbf{e}_{i}\cdot(\partial_{j}\varphi_{n}\mathbf{e}_{j})=s_{n}g_{ij}\partial_{i}(\partial_{j}\varphi_{n})=s_{n}g\partial_{ii}\varphi_{n}.
\label{eq4}
\end{equation}
Evidently the transformed equation is identical to the previous
one if we interpret the new equation as being in a right-handed
Cartesian system and keep $\theta$ and $\varphi_{n}$ unchanged.

The invariance of the eigenmodes $\varphi_{n}$ and their
corresponding eigenvalue $s_{n}$ immediately suggest a way to
studying the electrostatic response of the transformed structure.
Note that we can expand its total potential $\varphi_{t}$ in terms
of these eigenmodes,
\begin{equation}
\varphi_{t}(\omega)=\sum_{n}\frac{s(\omega)}{s(\omega)-s_{n}(1-ik^{2}/8)}(\varphi_{n}|\varphi_{0})\varphi_{n}\equiv\sum_{n}\beta_{n}(\varphi_{n}|\varphi_{0})\varphi_{n},
\label{eq5}
\end{equation}
where $k=\omega/c$, $s(\omega)=1/(1-\epsilon_{1})$ being a
spectral parameter and $\varphi_{0}$ being the external potential.
The radiation damping has been included by adding the factor
$ik^{2}/8$ \cite{draine,carminati}. Consequently, once these
eigenmodes of the original or seed geometry are known, we can
obtain the potential $\varphi_{t}$ of any transformed geometry by
just calculating the expansion coefficients
$(\varphi_{n}|\varphi_{0})$. In other words, transforming geometry
is equivalent to transforming the external source $\varphi_{0}$,
or more precisely, changing the expansion coefficients of each
eigenmodes. Furthermore, the time-averaged power absorbed by the
structure can be written as
\begin{equation}
P_{\textrm{a}}=\frac{\omega\epsilon_{0}}{2}\textrm{Im}(\epsilon_{1})\sum_{n}s_{n}|\beta_{n}(\varphi_{n}|\varphi_{0})|^{2}=\frac{\omega\epsilon_{0}}{2}\textrm{Im}(\epsilon_{1})I_{\textrm{e}},
\label{eq6}
\end{equation}
where $I_{e}$ represents the the integration of the electric field
intensity $|E|^{2}$ inside the particle. In a similar way, the
total power, the extinction, taken out by the particle has an
expression as
\begin{equation}
P_{\textrm{ex}}=\frac{\omega\epsilon_{0}}{2}\sum_{n}s_{n}|(\varphi_{n}|\varphi_{0})|^{2}\textrm{Im}[(\epsilon_{1}-1)\beta_{n}].
\label{eq7}
\end{equation}
When we transform the external field $\varphi_{0}$ accordingly,
i.e., $\varphi_{0}(x,y)=\varphi'_{0}(u,v)$ with $u$ and $v$ being
the new coordinates, all these energy quantities,
$P_{\textrm{a}}$, $I_{\textrm{e}}$ and $P_{\textrm{ex}}$, are
evidently invariant. On the other hand, if the external potential
is fixed in the $w$ coordinate with $w=u+iv$, we will find that
these energy quantities are proportional to $a^{2}$ when the
conformal mapping has a form as $a\times w(z)$ with $z=x+iy$ and
$a$ is real. In other words, bigger the particle area, stronger
the absorption and extinction.

It should be mentioned that the geometry resonances actually are
the \textit{surface modes} of the particle \cite{bohren,yong}, and
the resonance condition for the $n$th mode is strictly
$s(\omega)=s_{n}$, or equivalently $\epsilon_{1}=1-1/s_{n}$, when
we do not include the radiation loss. Notice that the
corresponding permittivity of the particle $\epsilon_{1}$ should
be real and negative since $0<s_{n}\leq 1$ \cite{bohren}. The most
striking property of the surface mode is that the resultant total
electrostatic energy of the whole system, including the free space
and the particle, is exactly zero \cite{wang}. Furthermore, the
complex frequency of the $n$th \textit{surface plasmonic
resonance} (SPR) is given by \cite{bergman,bergman2}
\begin{equation}
s(\omega_{n}-i\gamma_{n})=s_{n}, \label{eq8}
\end{equation}
with $\omega_{n}$ being real resonant frequency and $\gamma_{n}$
being the relaxation rate \cite{fredkin}. Since the eigenvalues
$s_{n}$ are conserved under any conformal transformation, the
transformed structure hence have same SPRs as the original
structure at the identical resonant frequencies. For instance,
since a metallic cylinder can be transformed from a
metal-dielectric interface by $w=e^{z}$, its SPR hence can be
determined by nonretarded surface-plasmon condition,
$\epsilon_{1}=-1$, of the metal-dielectric interface
\cite{zayatsa}.

\begin{figure}
\epsfxsize=300pt \epsfbox{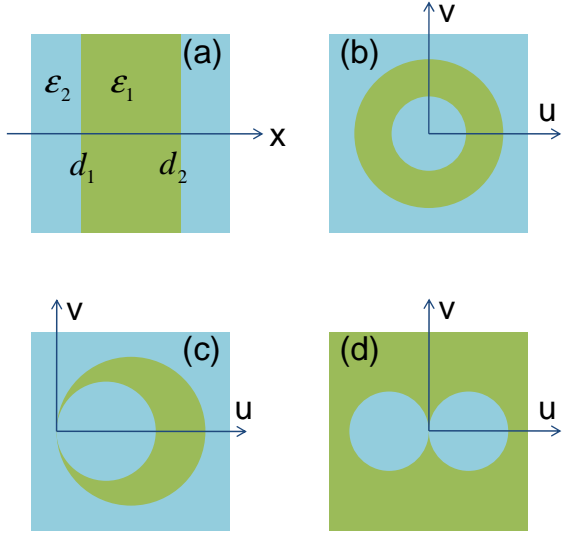} \vspace*{-5.5cm}
\caption{An one-dimensional finite-thickness slab in the $xy$
coordinates (a) is transformed to an annulus (b) with a conformal
transformation of $w=e^{z}$, a crescent (c) and two kissing
cylinders (d) with a conformal transformation of $w=1/z$, in the
new $uv$ coordinates. Here $w=u+iv$ and $z=x+iy$.} \label{fig1}
\end{figure}

As a proof of principle, we consider an one-dimensional dielectric
slab with finite thickness (Figure (1a)) and its derivative
systems obtained through different conformal transformations. The
eigenvalues of the slab are given by
\begin{equation}
s_{k,\pm}=\frac{1\pm e^{-|k|(d_{2}-d_{1})}}{2}, \label{eq9}
\end{equation}
where $k$ being real and nonzero, and the corresponding eigenmodes
are
\begin{equation}
\alpha_{k,\pm}\varphi_{k,\pm}e^{-iky} = \left\{
\begin{array}{lll}
            [e^{-2|k|d_{1}}\mp e^{-|k|(d_{1}+d_{2})}]e^{|k|x},  & x\leq d_{1} \\
            \mp e^{-|k|(d_{1}+d_{2})}e^{|k|x}+e^{-|k|x}, &d_{1}<x<d_{2} \\
            (1\mp e^{|k|(d_{2}-d_{1})})e^{-|k|x}, & x\geq d_{2} \\
\end{array} \right.
\label{eq10}
\end{equation}
where the normalize coefficient
$\alpha_{k,\pm}^{2}=8\pi|k|e^{-|k|d_{1}}(e^{-|k|d_{1}}\mp
e^{-|k|d_{2}})$. Evidently, in terms of the symmetry of electric
field, these eigenmodes can be cataloged into two groups. The
subscript $+(-)$ corresponds to even (odd) mode respectively,
while their eigenvalues are bigger (smaller) than 0.5. Since the
electric fields of the odd modes penetrate the metal weakly, they
can propagate longer along the metallic surface than the even
modes \cite{zayatsa}. Furthermore, the resonant condition,
approximately given by $\mathrm{Re}[s(\omega_{n})]=s_{n}$ when the
relaxation rate is weak, suggests that only the odd modes will be
excited resonantly when the real part of the dielectric
permittivity, $\textrm{Re}(\epsilon_{1})$, is smaller than $-1$.
Moreover, these eigenmodes and eigenvalues above can be directly
applied to the complementary structure of the finite slab, i.e. a
free-space gap sandwiched by two semi-infinite metallic spaces.
Since these two structures have identical geometry, they therefore
share same eigenvalues and eigenmodes. To obtain the induced
potential of the complementary structure, we only need to
interchange $\epsilon_{1}$ with $\epsilon_{2}$ in the definition
of $s(\omega)$. As a direct consequence, when
$\textrm{Re}(\epsilon_{1})$ is smaller than $-1$, only the even
modes will be resonantly excited in the semi-infinite structure
\cite{aubry6}.

We now calculate the resultant expansion coefficients of a
structure transformed from the finite slab through a conformal
mapping of $w=w(z)$. It is further assumed that the external
potential $\varphi_{0}(u,v)=p_{u}u+p_{v}v$, which corresponding a
uniform electric field $-p_{u}\mathbf{e}_{u}-p_{v}\mathbf{e}_{v}$.
The corresponding expansion coefficients can be written as
\begin{equation}
(\varphi_{k,\pm}|\varphi_{0})=\frac{2k}{s_{k,\pm}\alpha_{k,\pm}}\left[\varrho_{2}pF_{k}-\varrho_{1}
p^{\ast}F_{-k}\right], \label{eq11}
\end{equation}
where
\begin{equation}
F_{k}=\int_{d_{1}}^{d_{2}}dx\int_{-\infty}^{\infty}dy
\frac{dw}{dz}e^{kz^{\ast}},\:\:\: p=\frac{1}{2}(p_{u}-ip_{v}).
\label{eq12}
\end{equation}
In addition, $\varrho_{1}=1$ and $\varrho_{2}=\mp
e^{-|k|(d_{1}+d_{2})}$ for positive $k$, and they should exchange
when $k<0$.

To validate our approach, we first transforms the slab to two
coaxial cylinders (Figure (1b)) with $w=e^{z}$, and the radius of
the two cylinders are $r_{1}=e^{d_{1}}$ and radius
$r_{2}=e^{d_{2}}$ respectively. The expansion coefficients are
found to be
\begin{equation}
(\varphi_{k,\pm}|\varphi_{0})=\mp\frac{\alpha_{k,\pm}}{4}p_{u}e^{d_{1}+d_{2}}\delta(|k|-1),
\label{eq13}
\end{equation}
here $p_{v}$ is assumed to be zero because of the structural
symmetry. Evidently, only eigenmodes with $k=\pm 1$ are excited.
We further use these coefficients to calculate the induced
potential $\varphi_{i}$
\begin{equation}
\frac{\sigma\varphi_{i}}{\varphi_{0}} = \left\{
\begin{array}{lll}
            (r^{2}_{1}-r^{2}_{2})(1-\epsilon_{2})^{2},  & r\leq r_{1} \\
            r^{2}_{1}(1-\epsilon_{2})^{2}+r^{2}_{2}(1-\epsilon_{2}^{2})-2(1-\epsilon_{2})r^{2}_{1}r^{2}_{2}/r^{2}, &r_{1}<r<r_{2} \\
            r_{2}^{2}(r^{2}_{2}-r^{2}_{1})(1-\epsilon_{2}^{2})/r^{2}, & r\geq r_{2} \\
\end{array} \right.
\label{eq14}
\end{equation}
with
$\sigma=r^{2}_{2}(1+\epsilon_{2})^{2}-r^{2}_{1}(1-\epsilon_{2})^{2}$.
Our results are exactly identical to the one obtained by expanding
the potential with Bessel functions \cite{bohren}. Furthermore, by
letting $r_{1}\rightarrow 0$ or $r_{2}\rightarrow\infty$, the
results above can be used to describe a dielectric cylinder of a
cylinder void \cite{bohren}.

Since the eigenvalues of the finite slab cover a wide region
$[0,1]$ (note that the eigenvalues of the eigenproblem (\ref{eq1})
are limited between 0 and 1), any negative
$\textrm{Re}(\epsilon_{1})$ will excite a surface mode as long as
its expansion coefficient is not zero. Consequently the
transformed system will present broadband response in principle.
One example is the crescent studied in Ref. \cite{aubry1}, which
can be obtained by using $w=1/z$ when $d_{1}>0$ (Figure (1c)). The
expansion coefficients are found to be
 \begin{equation}
(\varphi_{k,\pm}|\varphi_{0})=\frac{\alpha_{k,\pm}}{4}\left[p_{u}+i\textrm{sgn}(k)p_{v}\right].
\label{eq15}
\end{equation}
Evidently, each eigenmode is excited by the external potential.
Furthermore, $|(\varphi_{k,\pm}|\varphi_{0})|$ depends on the
amplitude of the external electric field exclusively. The energy
quantities such as the absorption hence do not depend on the
direction of the incident field \cite{aubry1}, a property which is
not so obviously. Another example is the kissing cylinders
suggested in Ref. \cite{aubry6}, which is obtain by transforming
two semi-infinite slabs, with $d_{1}<0$ and $d_{2}>0$, with
$w=1/z$ (Figure (1d)). The corresponding expansion coefficients
are calculated as
 \begin{equation}
(\varphi_{k,\pm}|\varphi_{0})=\frac{\alpha_{k,\pm}}{4}\left[p_{u}\frac{e^{|k|(d_{2}+d_{1})}\pm
1}{e^{|k|(d_{2}-d_{1})}\pm
1}+i\textrm{sgn}(k)p_{v}\frac{e^{|k|(d_{2}+d_{1})}\mp
1}{e^{|k|(d_{2}-d_{1})}\mp 1}\right]. \label{eq16}
\end{equation}
Again all the eigenmodes can be excited by the external source.
Note that when $d_{1}=-d_{2}$ the cylinders have same radius. The
kissing cylinders then possess both $u$ and $v$ mirror symmetry.
Consequently, the even mode $\varphi_{k,+}$ or the odd mode
$\varphi_{k,-}$ can be excited by $u$ or $v$- polarized electric
field alone \cite{aubry3,aubry5}.

As suggested by the equation (\ref{eq6}), the integration of the
electric field intensity $I_{e}$ inside the particle is invariant
under conformal transformation. On the other hand, the area $A$ of
the particle is generally changed significantly. As a direct
result, the degree of the local field localization, approximately
measured by $I_{e}/A$, will be modified by the transformation.
Taking the kissing cylinders or the crescent above as an example.
Before transformation, the area of the one-dimensional slab is
infinite. The transformed structure, on the other hand, has a
finite area. Consequently, the kissing cylinders or the crescent
strongly enhances the local field
\cite{aubry1,aubry2,aubry3,aubry4,aubry5,aubry6}.

A few words regarding three-dimensional transformation optics
techniques in electrostatic, which has been developed in Ref.
\cite{dominguez} lately. Under a general transformation, the
three-dimensional eigen-equation, equation (\ref{eq1}), will be no
longer invariant and the resultant expression usually is very
complicated. One exception is that the Jacobian matrix
$\mathcal{J}$ of the transformation satisfies
$\mathcal{J}\mathcal{J}^{T}=f^{-1}(r')\mathcal{I}$, with
$\mathcal{I}$ being the unit $3\times 3$ matrix and $f(r')$ being
an arbitrary function of $r'$ (the transformation used in Ref.
\cite{dominguez} possesses this property). Under this condition,
the new eigenmodes $\phi_{n}$ connect to the previous eigenmodes
$\varphi_{n}$ of the original system as
\begin{equation}
\nabla\phi_{n}=\sqrt{f}\nabla\varphi_{n}. \label{eq17}
\end{equation}
The corresponding eigenvalues $t_{n}$ are also different with the
previous $s_{n}$
\begin{equation}
t_{n}=\frac{\int\theta f\nabla\varphi_{n}dv}{\int
f\nabla\varphi_{n}dv},\:\:\:s_{n}=\frac{\int\theta
\nabla\varphi_{n}dv}{\int\nabla\varphi_{n}dv}, \label{eq18}
\end{equation}
when the integrations are performed over the previous system. Once
the new eigenmodes and eigenvalues are obtained, equation (5-7)
can be directly employed without any modification. Note that for a
two-dimensional conformal transformation equation (17,18) still
work, by simply replacing the function $f$ with a constant number.
Again, we find that the eigenmodes and eigenvalues are invariant
under any two-dimensional conformal transformation.

In conclusion, we proved that the electrostatic eigenmodes and
eigenvalues of two-dimensional nanosystems are invariant under any
conformal transformation. Based on this property, we suggested an
new approach to studying the electrostatic responses of the
transformed structures. Namely, transforming a geometry is
equivalent to transforming the external potential or the expansion
coefficients of the invariant eigenmodes. Furthermore, except the
potential, energy quantities such as absorption, extinction and
electrostatic energy are also found to be conserved. Moreover, the
significant conditions to designing broadband nanosystems are the
eigenvalues of the system should cover a wide portion of $[0,1]$
as well as the external source should excite nearly all the
eigenmodes.

This work was supported in part by the Penn State MRSEC under NSF
grant no. DMR 0213623.

\end{document}